\documentclass[singlecol,noshowpacs]{epl2} 
\usepackage[margin=2cm]{geometry}
\usepackage{amssymb}
\usepackage{mathrsfs}
\usepackage{graphics}
\usepackage{epsfig}
\usepackage{graphicx}
\usepackage{subfigure}
\usepackage{bm}

\def\beq{\begin{equation}}
\def\eeq{\end{equation}}
\def\baq{\begin{eqnarray}}
\def\eaq{\end{eqnarray}}

\def\fnl{f_{\rm NL}}

\def\gnl{g_{\rm NL}}

\def\taunl{\tau_{\rm NL}}
\def\hnl{h_{\rm NL}}

\def\x{{\bf x}}

\def\fnl{f_{\rm NL}}

\def\gnl{g_{\rm NL}}

\def\bea{\begin{eqnarray}}
\def\eea{\end{eqnarray}}
\def\be{\begin{equation}}
\def\ee{\end{equation}}

\def\Pz{{\mathcal P}_{\zeta}}

\def\fnlo{f_{\rm NL}^{\rm obs}}

\def\gnlo{g_{\rm NL}^{\rm obs}}

\def\bnl{\beta_{\rm NL}}

\def\vk{\vec{k}}

\def\Nin{N_{\rm in}}

\def\f{f_{\rm NL}^{\rm obs}}
\def\fo{f_{\rm NL}^{0}}
\def\s{\sigma}

\def\G{{\rm G}}
\def\l{{\rm l}}
\def\sf{\sigma_{\fnl}}

\begin{document}
\title{Implications of the Planck bispectrum constraints for the primordial trispectrum}

\author{Christian T.~Byrnes,$^{\,a}$
  Sami Nurmi,$^{\,b\,}$
  Gianmassimo Tasinato,$^{\,c\,}$
 David Wands$^{\,c\,}$}

\institute{$^a$Astronomy Centre, University of Sussex, Brighton, BN1 9QH, UK, \\ $^b$Department of Physics and Helsinki Institute of Physics, University
of Helsinki, P.O. Box 64, FIN-00014 University of Helsinki, Finland
\\
$^c$Institute of Cosmology $\&$ Gravitation, University of Portsmouth, Dennis Sciama Building \\ \hskip0.2cm Portsmouth, PO1 3FX, United Kingdom }

\abstract{ 
The new Planck constraints on the local bispectrum parameter $\fnl$ are about $10^5$ times tighter than the current constraints on the trispectrum parameter $\gnl$, which means that the allowed numerical values of the second and third order terms in the perturbative expansion of the curvature perturbation are comparable. We show that a consequence of this is that if $\gnl$ is large enough to be detectable, then it will induce a large variation between the observable value of $\fnl$ and its value in a larger inflated volume. Even if there were only a few extra efoldings between the beginning of inflation and horizon crossing of our Hubble horizon, an observably large $\gnl$ means that $\fnl$ is unlikely to be as small as its current constraint, regardless of its true background value. This result is very general, it holds regardless of how many fields contributed to the curvature perturbation. We also generalise this result to other shapes of non-Gaussianity, beyond the local model. We show that the variance of the 3-point function in the squeezed limit  is bounded from below by the square of the squeezed limit of the 4-point function.} 


\maketitle

\section{Introduction}\label{intro}

One of the most keenly anticipated results from the Planck 2013 data release was the measurement of the primordial bispectrum \cite{Ade:2013ydc}. Constraints on specific forms of non-Gaussianity, parameterised by the non-linearity parameter $\fnl$, have significantly improved compared to the previous constraints \cite{Bennett:2012fp,Giannantonio:2013uqa}.
Given that the bispectrum is only the first in a hierarchy of correlators that parameterise deviations from a Gaussian distribution, it is interesting to ask whether these tight constraints on $\fnl$ also have implications for the likely value of higher-order correlators in general. 
In this letter we show that measurements of $\fnl$ do provide information about the likely value of the trispectrum 
 in inflationary models assuming only that a nearly scale-invariant spectrum of fluctuations exists on scales slightly larger than our observable horizon today.

We start with the local Ansatz for the primordial curvature perturbation defined in a large reference volume (denoted by $\,^0$) which may be larger than the observable universe
\bea
 \label{zetalocal}
 \zeta=\zeta_G+\frac35\fnl^0\left( \zeta_G^2 - \langle\zeta_G^2\rangle \right) +\frac{9}{25}\gnl^0\zeta_G^3,
  \eea
where $\zeta_G$ is a Gaussian random field.
The current constraints on the non-linearity parameters (in our observable patch, denoted by $\,^{\rm obs}$) are \cite{Ade:2013ydc,Sekiguchi:2013hza}
\bea \fnlo&=&2.7\pm5.8, \\
\gnlo&=&(-3.3\pm2.2)\times10^5.
\eea
Note that no $\gnl$ constraint has yet been made with Planck data (the given constraint is from WMAP9 data \cite{Sekiguchi:2013hza}). The forecast $1\sigma$ error bar for Planck data has been estimated as $\sigma_{\gnl}=6.7\times10^4$ by \cite{Sekiguchi:2013hza} and $\sigma_{\gnl}\simeq1.3\times10^5$ by \cite{Smidt:2010ra}. In order to have a clear detection of $\gnl$ with Planck, we require
\bea
 \label{gobs}
 |\gnlo|\gtrsim2\times10^5 \,.
 \eea
This means that the third order term in (\ref{zetalocal}) is not more tightly constrained than the second order term. The question is whether there exist values of $\fnl^0$ and $\gnl^0$
for which we are likely to have an observable
$|\gnlo|>2\times10^5$, while respecting the observational bound,
$|\fnlo|\leq15$.

Although $\fnl^0$ represents the most likely value of the $\fnl$ parameter measured by Planck in our observable patch, there is a natural variance in the actual value of $\fnl$ observed in any patch of the larger reference volume, and in inflationary models for the origin of structure, the variance of the observed $\fnl$ grows with the trispectrum parameter $\gnl$ and higher-order correlators \cite{Byrnes:2011ri,Verde:2013gv}.

To demonstrate this we split the first-order (Gaussian) curvature perturbation into long and short wavelength parts, $\zeta_G=\zeta_{G,l}+\zeta_{G,s}$, where the splitting scale is defined by the horizon scale today, so that the short wavelength modes are those which we observe, while the long wavelength mode modulates the background value in our observable patch. The observed curvature perturbation in our patch is given by
\bea
 \zeta_s=\zeta_{G,s}+\frac35\fnlo\left( \zeta_{G,s}^2 - \langle\zeta_{G,s}^2\rangle \right)+\frac{9}{25}\gnlo\zeta_{G,s}^3 \,,
 \eea
where \cite{Byrnes:2011ri}
 \beq
  \label{fnlobs_exp}
 \fnlo=\fnl^0+\left(\frac95\gnl^0-\frac{12}{5}(\fnl^0)^2\right)\zeta_{G,l}+ {\cal O}(\hnl^0\zeta^2_{G,l})\
 ,
 \qquad \gnlo=\gnl^0+ {\cal O}(\hnl^0\zeta_{G,l})\ .
 \eeq
We assume that the coefficient $\hnl^0$ of the
fourth-order term in the expansion (\ref{zetalocal}) is not
extremely large $h_{\rm NL}^0\lesssim \gnl^0/\zeta_{G,l}$, and
similarly for the higher-order terms. The long-wavelength modes then
do not change the order of magnitude of $\gnl^{\rm obs}$ and we can
set $\gnl^{\rm obs}\simeq \gnl^0$. However, the variation of
$\fnl^{\rm obs}$ can be significant if $|\gnl^0|\gg (\fnl^0)^2$.

The long-wavelength part of the curvature perturbation,
$\zeta_{G,l}$, is due to 
fluctuations on scales outside our observable patch. This may be due to $N_{\rm
in}$ e-foldings of inflation before our observable scale left the Hubble-horizon during inflation.
We may therefore estimate the typical long wavelength curvature fluctuation
 \bea \zeta_{G,l}
  \label{zetaGl}
 \approx \sqrt{\Pz N_{\rm in}}. \eea
Using $\sqrt{\Pz}\simeq5\times10^{-5}$, and taking (\ref{gobs}) and (\ref{fnlobs_exp}) yields that an observably large $\gnl$ requires that on average
 \bea
 \label{variance}
 |\fnlo-\fnl^0|\gtrsim20 \sqrt{N_{\rm in}}\ .
\eea  
We can see that observably large $\gnl^{\rm obs}$ makes $\fnl^{\rm
obs}$ likely to vary from its background value $\fnl^0$ by more than
its observed upper limit (and more than the Planck error bar), even if there is only one e-folding before
observable scales exited the horizon, $N_{\rm in}=1$. 
Hence there appears to be a tension between having a
small $\fnl^{\rm obs}$ together with a value $\gnl^{\rm obs}$
large enough to be observable.
More generally, an extended period of inflation may lead to a large variance in the locally observable bispectrum parameter, $\fnlo$. The variance is proportional to the trispectrum parameter, $\gnl^2$, and the duration of inflation. 

A local distribution for the primordial curvature perturbation (\ref{zetalocal}) arises naturally when the cosmological expansion on super-Hubble scales, $N=\int H dt$, is a local function of a Gaussian distribution of scalar field perturbations during inflation, $\delta\varphi$ \cite{Lyth:2005fi},
\be
 \label{deltaN}
 \zeta = N' \delta\varphi + \frac12 N'' \delta\varphi^2 + \frac16 N''' \delta\varphi^3+\cdots \,,
\ee
where we identify $\zeta_G=N'\delta\varphi$ and
\be
 \fnl^0 = \frac56 \frac{N''}{N^{\prime2}} \,,
 \quad
 \gnl^0 = \frac{25}{54}\frac{N'''}{N^{\prime3}} \,.
 \ee
Of course Eq.~(\ref{zetalocal}) only represents the first three terms in a Taylor series expansion, but it is sufficient to illustrate the general principle.

Assuming $N_{\rm in}$ e-foldings of inflation from the start of inflation before our
observable patch left the Hubble-horizon, we have $\langle\delta\varphi_l^2\rangle\simeq {\cal P}_\varphi N_{\rm in}$ for a massless scalar field, and hence we obtain the large-scale curvature perturbation, (\ref{zetaGl}).
If the scalar field has a finite and positive effective mass-squared ($m^2>0$) during inflation then after many e-folds of inflation the variance of the field reaches an equilibrium value~\cite{Bunch:1978yq,Starobinsky:1994bd}
\be
 \langle\delta\varphi_l^2\rangle \simeq \frac{3}{8\pi^2}\frac{H^4}{m^2} \,,
\ee
equivalent to a limiting value $N_{\rm in}\to (2\eta)^{-1}$ in Eq.~(\ref{zetaGl}) where the slow-roll parameter $\eta\equiv m^2/3H^2\ll1$. In general, we may expect corrections to our assumption of a scale invariant spectrum for $\Nin\gtrsim1/(n_s-1)$, as discussed in this context in \cite{LoVerde:2013xka}.

Our result~(\ref{variance}) differs to the arguments that most known models predict either $\gnl\simeq\fnl$ or $\gnl\simeq\fnl^2$ \cite{Suyama:2013nva}. Those arguments give a tighter bound on $\gnl$, but are
model dependent and derived on a case by case basis, whilst ours is almost completely model independent. Our only assumption is that the power spectrum remains almost scale invariant over a range of scales slightly larger than the observable horizon today.

\section{The variance of $\fnlo$ from $\gnl$} \label{gnl-constraint}

The probability distribution for the observed value of $\fnl^{\rm
obs}$ is determined by the expansion (\ref{fnlobs_exp}) and the
statistics of the long-wavelength fluctuations $\zeta_{\G,\l}$ over
the inflating volume. $\zeta_{\G,\l}$ is Gaussian
and we truncate the series (\ref{fnlobs_exp}) to first order. The
probability distribution for $\fnl^{\rm obs}$, given a mean
value $\fo$ over the inflating volume, is given by  \cite{Nurmi:2013xv}
\beq
\label{Pf}
P(\f|\sf,\fo) = \frac{1}{\sqrt{2\pi}\sf} \exp\left(
-\frac{(\f-\fo)^2}{2\sf^2} \right) \ .
\eeq
The variance is determined by the global mean values $\fnl^0$ and
$\gnl^0$ and the amount of inflation $\Nin$ before the
horizon crossing of our observable patch, see (\ref{fnlobs_exp}),
  \beq
  \label{sigmafnl-first}
  \sigma_{\fnl}^2 = \left(\frac{9}{5}\gnl^0-\frac{12}{5}(\fnl^0)^2\right)^2{\cal
  P}_\zeta \Nin\ .
  \eeq
Here ${\cal P}_\zeta$ measures the spectrum of long-wavelength modes.
For $|\fnl^0|\lesssim 100$ and $\Nin\lesssim 100$, this
coincides with the observable spectrum, 
${\cal P}^{\rm obs}\simeq{\cal P}^{0}$ \cite{Nurmi:2013xv}. 

\begin{figure}[h!]
    \begin{center}
    \includegraphics[width=9 cm, height= 5 cm]{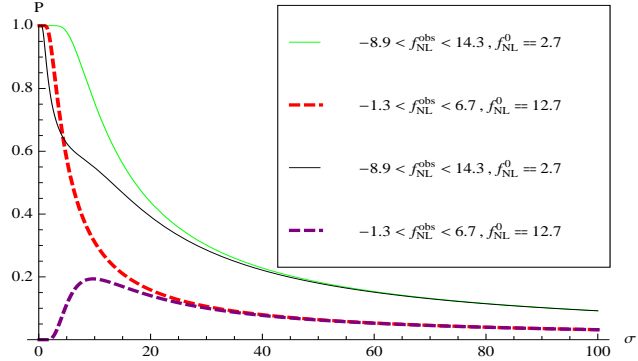}
    \caption{The probability $P(\fnl^{\rm min}<\fnlo<\fnl^{\rm max}|\s,\fo)$. We show two choices of the allowed range for $\fnlo$, firstly the $2\sigma$ Planck constraints (solid lines) and secondly a future experiment with the same central value, $2.7$, and an allowed  range of $\pm4$ (dashed lines), each case with two
choices of $\fnl^0$ as denoted in the figure legend.}
    \label{fig:fnl-pdf}
    \end{center}
  \end{figure}

As the variance $\sigma_{\fnl}$ grows, the distribution flattens out
and the probability to find $\fnl^{\rm obs}$ close to the global
mean $\fnl^0$ decreases. This behaviour is illustrated in
Fig.~\ref{fig:fnl-pdf} which shows the probability of finding
$\fnl^{\rm obs}$ in the Planck $2\sigma$ interval as a function
of $\sigma_{\fnl}$ and for two fixed choices of the global mean
$\fnl^{0}$. We also plot the analogous results for a hypothetical experiment with a $2\sigma$ range of $\pm4$ and the same mean value.
The probability for $\sigma=0$ is either unity or zero, depending on whether $\fnl^0$ lies within the allowed range for $\fnlo$. The figure shows that the probability for $\fnl^{\rm obs}$
in our patch being small gets significantly suppressed for
$\sigma_{\fnl}\gtrsim 10$. This is so even if the underlying
inflationary model would predict $\fnl^0=2.7$ over the entire inflating
volume. As (\ref{sigmafnl-first}) is controlled by the
magnitude of the trispectrum, inflationary models
with a large $\gnl^0$ are less likely to generate a small bispectrum
in our observable patch.

\begin{figure}[h!]
  \centering
    \includegraphics[width=15 cm, height= 5.5 cm]{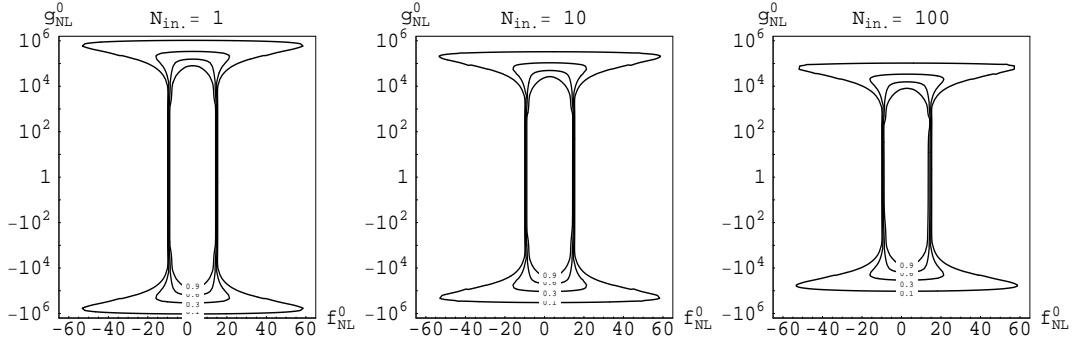}
  \caption{Contour plots showing probability for finding $\fnl^{\rm obs}$ within the
  $2\sigma$ bounds of Planck, $-8.9<\fnl^{\rm obs}<14.3$ as a function of $\fnl^0$ and $\gnl^0$ for three values of $N_{\rm in}$.
   }
  \label{fig:fnlnondet}
\end{figure}

This is shown  in Fig.~\ref{fig:fnlnondet}, which depicts
the probability that $-8.9<\fnl^{\rm obs}<14.3$, corresponding to
the 95\% C.L. bounds of Planck, as a function of the global mean
values $\fnl^0,\;\gnl^0$, and the number of e-foldings before
our horizon crossing.
We observe that for large trispectrum amplitudes, $|\gnl^0|\gtrsim
10^5$, the probability of obtaining a small $\fnl^{\rm obs}$
compatible with the observational bounds drops below $30$\% even for
$N_{\rm in}\sim 1$ and for $\fnl^0=0$. If there was at least $\Nin\sim 100$ e-foldings of inflation before our horizon exit, the
probability is further suppressed to below the $10$\% level.

We have seen that a large value of $\gnl^{\rm obs}$ makes a small
value of $\fnlo$ less likely. We may also 
 ask whether a small value of $\fnlo$ makes a large value of
$\gnl^{\rm obs}$ unlikely. Using Bayes' theorem we can write
down an expression for the probability of the variance
$\sigma_{\fnl}$ given a mean $\fo$ and observed value $\fnl^{\rm
obs}$, $
P(\s|\f,\fo) = P(\f|\s,\fo)P(\s,\fo)/P(\f,\fo)$.
Using this, we could work out the constraints on $\sf$, and hence
$\gnl^0$. However, due to
the slow convergence of the distribution $P(\fnl^{\rm
obs}|\sf,\fo)$, which is manifest as the tail in
Fig.~\ref{fig:fnl-pdf}, the constraints turn out to be strongly
dependent on the choice of prior for the variance, $\sigma_{\rm max}$.
With a physically motivated
prior choice $\sigma_{\rm max}$, or correspondingly a prior upper
limit for $|\gnl^0|$, one could obtain useful model dependent
a posteriori constraints on $\gnl^0$. Here we however wish to keep our
analysis model independent. We therefore conclude that we cannot
obtain a posteriori bounds on $\gnl^0$ given only constraints on $\fnlo$.

\section{The variance of $\fnlo$ is larger in multi field scenarios}\label{multi-sec}

The preceding results can easily be generalised to multi-field models where
the superhorizon scale curvature perturbation is generated by more than
one scalar field. 
Assuming again that the field fluctuations are Gaussian,
the curvature perturbation over the entire inflating patch can be expanded as
  \beq
  \label{zeta_multi}
  \zeta(x)=\sum_{a=1}^{n}N_a\delta\sigma_a(\x)+\frac12\sum_{a,b=1}^{n}N_{ab}\delta\sigma_a(\x)\delta\sigma_b(\x)+\cdots\
  ,
  \eeq
where $n$ denotes the number of scalar fields and henceforth we assume summation over repeated indices. This coincides with the usual $\delta N$ expression \cite{starob85} provided that the
fluctuations $\delta\sigma_{a}$ are evaluated on a spatially flat
initial hypersurface, and $N(\sigma)$ then measures the number of
e-foldings. However for our purposes it is not necessary to make
this identification and we can understand equation
(\ref{zeta_multi}) as a generic Taylor expansion.

Proceeding analogously to the single-field case \cite{Nelson:2012sb,Nurmi:2013xv},
we expand the locally observable bispectrum amplitude to first-order
in long-wavelength fluctuations as
  \beq
  \fnl^{\rm
  obs}(\x_0)=\fnl^{0} + \sum_{a=1}^{n}\frac{\partial\fnl^0}{\partial\sigma_{a}}\,\delta_{\rm
  l}\sigma_{a}(\x_0)+\cdots\ .
  \eeq
Here $\fnl^0=(5/6)N_{a}N_{b}N_{ab}/(N_cN_c)^2$ denotes the
tree-level bispectrum amplitude averaged over the entire inflating
patch and $\x_0$ labels the location of our observable patch.

The variance of $\fnl^{\rm obs}$ is then given by the expression
  \baq
  \label{sigmafnl}
  \sigma_{\fnl}^2&=&f_{{\rm NL},a}f_{{\rm NL},b}\langle\delta_{\rm L}\sigma_a\delta_{\rm L}\sigma_b\rangle
  \\\nonumber&=&\frac{25}{36} \left(4 \tau_6^{(1)}+
\tau_6^{(2)}+\frac{576 \fnl^{2}}{25}\,  \tau_{\rm NL} +4
g^{(1)}_6-\frac{96 \fnl }{5} \,f_5^{(1)} -\frac{48  \fnl }{5}
f_5^{(2)} \right)\,{\cal P}_\zeta \Nin,
  \eaq
where the quantities on the right hand side denote global background
values but we have omitted the labels ``${0}$'' for brevity. Here we
have defined \cite{Alabidi:2005qi,Suyama:2011qi}
  \baq
   &\taunl=\frac{N_a N_{ab}N_{bc}N_c}{\left(N_d N_d\right)^3}, \;\;f_5^{(1)}=\frac{N_a N_{a b} N_{b c} N_{c d } N_d}{\left( N_e N_e \right)^4},\;\;
 f_5^{(2)}=\frac{N_a N_{a b} N_{b c d} N_{c } N_d}{\left( N_e N_e \right)^4}, \\ &\tau_6^{(1)}= \frac{ N_a N_{a b} N_{b c } N_{c d } N_{d e} N_e
   }{\left( N_f N_f \right)^5},\;\;  \tau_6^{(2)}= \frac{ N_a
  N_b N_{a b c} N_{ c d e } N_{d } N_e }{\left( N_f N_f \right)^5},\;\; g_6^{(1)} =\frac{ N_a N_b N_{a b c} N_{ c d  }
  N_{d e } N_e }{\left( N_f N_f \right)^5}.  \nonumber
  \eaq

Using the Cauchy-Schwarz inequality $(f_{{\rm
NL},a}N_a)^2\leq(f_{{\rm NL},a}f_{{\rm NL},a})(N_bN_b)$ we find that
the variance is bounded from below by
  \beq
  \sigma_{\fnl}^2\geq\left(\frac{9}{5}\gnl+\frac{5}{3}\taunl-\frac{24}{5}\fnl^2\right)^2{\cal
  P}_{\zeta}N_{\rm in},
  \label{sigmafnl-multi}\eeq
 where \cite{Byrnes:2006vq}
 \beq
 \gnl = \frac{25}{54} \frac{N_aN_bN_cN_{abc}}{(N_dN_d)^3}.
 \eeq
In the single-field limit the inequality becomes saturated and reads
  \beq
  \sigma_{\fnl}^2=\left(\frac{9}{5}\gnl-\frac{12}{5}\fnl^2\right)^2{\cal
  P}_{\zeta}N_{\rm in},
  \eeq
as found in \cite{Nurmi:2013xv}. The quantities on the right hand
sides of these equations again denote the global background values.

We thus find the general result that in models where the ensemble
expectation value of the trispectrum amplitude $\gnl^0$ is much larger
than the bispectrum amplitude, $|\gnl^0|\gg (\fnl^0)^2$, the variance of
the locally observable bispectrum amplitude $\fnl^{\rm obs}$ is
bounded from below by
  \beq
  \label{sigmafnl_largegnl_ineq}
  \sigma^2_{\fnl}\geq\left(\frac{9}{5}\gnl^0\right)^2{\cal
  P}_{\zeta}N_{\rm in}.
  \eeq
The global amplitude $\gnl^0$ corresponds to
that of the locally observable trispectrum amplitude $\gnl^{\rm
obs}$ provided that the amplitudes of the higher-order
connected correlators are not extremely large.

We conclude that the variance of $\fnlo$ in single-source models, (\ref{sigmafnl-first}), is the most conservative, the presence of multiple sources generating $\zeta$ increases the variance.

We briefly consider how the amplitude of $\taunl$ affects the variance of $\fnl$. $\taunl$ may be observable with Planck or other data in the near future if $\taunl\gtrsim10^3$, the current Planck constraint is $\taunl\leq2800$ \cite{Ade:2013ydc}. Combined with the observational bound $|\fnlo|\lesssim10$, this requires $\taunl/\fnl^2\gtrsim10$, which is far from the single--source equality, $\taunl=(6\fnl/5)^2$, and is hard to realise for known models, e.g.~\cite{Peterson:2010mv,Elliston:2012wm,Leung:2013rza}, but can be constructed with fine tuning, e.g.~\cite{Ichikawa:2008ne,Byrnes:2008zy}. We have found that for two-field models, the relative variances are given by 
\bea \frac{\sigma_{\fnl}}{\fnl}\simeq\frac{\sigma_{\taunl}}{\taunl}\simeq  10^{-4}\frac{\taunl^{3/2}}{\fnl^2}\sqrt{r_T \Nin}, \eea
where $r_T$ is the tensor-to-scalar ratio. Although this quantity can be made large, typically this is only the case when both $\fnl$ and $\taunl$ are too small to be observable. In summary, an observably large $\gnl$ will imply a large uncertainty in relating the background and observed values of $\fnl$, but an observably large $\taunl$ generically does not.

\section{Extensions beyond local non-Gaussianity}

In the previous sections, we have
used the fact that the variance of the $\fnl$ parameter  of local non-Gaussianity is bounded from below by a quantity depending on the local trispectrum
parameter $\gnl$, as well as on $\taunl$ and $\fnl$.
We now
 discuss a
similar bound for the variance of a generalised $\fnl$ defined in the squeezed limit, without specifying any
particular shape of non-Gaussianity. In order to do so, we
consider an appropriate extension of the analysis of
\cite{Assassi:2012zq} (see also \cite{Smith:2011if,Kehagias:2012pd}) to higher-point functions, defining a new
inequality between non-Gaussian parameters that holds for
arbitrary shapes of non-Gaussianity.

In full generality,
 the parameter $\fnl$ 
is defined in the squeezed limit as \be \label{defnl}\lim_{\vk_1\to0}\langle
 \zeta_{\vk_1}
 \zeta_{\vk_2} \zeta_{\vk_3}
 \rangle_c'\,=\,\frac{12}{5}\,\fnl\,P_{\zeta}(k_1) P_{\zeta}(k_2).
\ee This quantity quantifies   how much the 2-point (pt) function of
the curvature perturbation, $\langle \zeta_{\vk_2} \zeta_{\vk_3}\rangle$,  is modulated by a long wavelength mode $k_1\to0$.
 We use the same conventions as \cite{Smith:2011if}: the
notation $\langle \dots \rangle_c'$ denotes the connected $n$-pt function
without the multiplicative factor $(2\pi)^3 \delta(\sum \vec{k})$.
We are interested in the variance of the parameter $\fnl$: this variance depends 
on the variances of the bispectrum $\langle
 \zeta_{\vk_1}
 \zeta_{\vk_2} \zeta_{\vk_3}
 \rangle_c'$, and of the power spectra $P(k_i)$. The variance of the power spectrum  is 
proportional to the trispectrum parameter $\taunl$ \cite{Byrnes:2011ri}.   This variance is observationally constrained, and for brevity, we will neglect it in this section.  We instead
focus  on the potentially large contribution to the variance of the bispectrum that is associated with the
trispectrum parameters. In other words, using
(\ref{defnl}), 
 we express the variance of $\fnl$ in the squeezed
limit as
\be
\sigma_{\fnl}\,=\,\frac{5}{12}\,\frac{\sigma_3}{P_{\zeta}(k_1) P_\zeta(k_2)},
\ee
where $\sigma_3$ is the variance of the bispectrum (in the same squeezed limit),
and $P_{\zeta}(k_1)P_\zeta(k_2)$ are the power spectra appearing in (\ref{defnl}).

In order to calculate the variance $\sigma_3$, we define $\alpha_3$ by the following Fourier transform
\be\label{defal}
\alpha_3^2(\vec{k})\,=\,\int \,d^3 x\,e^{i \vec{k} \vec{x}}\, \left[\langle \zeta^3(x)
\,
\zeta^3(0)\rangle'-\langle \zeta^3(x) \rangle' \langle \zeta^3(0) \rangle'
\right].
\ee
We write $\alpha_3$ as a function of the internal momentum $\vec{k}$ connecting
two 3-pt functions that form a 6-pt function.
The second term inside the square parenthesis of the previous expression, 
although generally needed for defining a variance, is nevertheless proportional
to the square of 3-pt functions, i.e.~$\fnl^2$. We will neglect its contribution in this section since the bounds on $\fnl$ constrain this contribution to be too small to be of interest.

In order to complete our definition of the variance, we consider the role 
of soft, long wavelength modes that connect the two three-point functions that appear
in $\langle \zeta^3(x)
\,
\zeta^3(0)\rangle'$.   
   Soft modes of small momenta $q\ll k$ that connect the 3-pt functions should be
 summed up  in the definition of the variance. Following 
 the arguments we developed in \cite{Tasinato:2012js}, 
  they cannot be individually measured and hence their
  effect must be taken into account by summing over them. Hence we adopt the following
  definition for the effective variance $\sigma_{3}$ of 3-pt functions in momentum 
  space:
  \be\label{intsm}
  \sigma^2_{3}(k)\,=\,\int_{k_L}^k\,\frac{d^3q}{(2 \pi)^3}\,\alpha^2_3(q)
  \ee  
 where $k_L$ is an infrared cut-off.  Following the arguments
 of \cite{Assassi:2012zq} (generalized to higher-point functions) one may show that 
$\alpha_3^2(\vec{q})$, 
 the Fourier transform of $
\langle \zeta^3(x)
\,
\zeta^3(0)\rangle'$ evaluated at the scale $\vec{q}$,   satisfies the following
inequality

\be\label{ineq1}
\alpha_3^2({\vec{q}})
\,\ge\,
\frac{| \langle \zeta({\vec{q}})
\,
\zeta^3(0)\rangle' |^2}{P_{\zeta} (q)}.
\ee
Taking the soft limit $\vec{q}\to0$, we get a general inequality relating the 
collapsed limit of a 6-pt function with the squeezed limit of a 4-pt function,
valid for arbitrary shapes. It is convenient to  re-express the quantity in the numerator
on the right hand side of (\ref{ineq1}) in terms of  a new non-Gaussian parameter $\bnl$,
defined in the squeezed limit as
\be \bnl\,\equiv\,\lim_{\vec{k}_1\to 0}\,\frac{\langle \zeta_{\vk_1}
 \zeta_{\vk_2} \zeta_{\vk_3} \zeta_{\vk_4}
 \rangle_c'}{P_{\zeta}(k_1) P_{\zeta}(k_2) P_{\zeta}(k_3) }\,.
\ee
At least in non-Gaussian models similar to the local model, $\bnl$ has the advantage of being almost scale independent. At this point, we integrate both sides of the inequality (\ref{ineq1}) along the soft mode $\vec{q}$.
 Repeating steps similar to \cite{Assassi:2012zq}, we find the inequality
 \be
 \int_{\vec{q}_1} \int_{\vec{q}_2} \int_{\vec{q}_3} \int_{\vec{q}_4}\,P_{\zeta} (q_1) \,
 P_{\zeta} (q_2) \, P_{\zeta} (q_3) \, P_{\zeta} (q_4) \,
 \left[ 
  \sigma_{\fnl}^2\,-\,\left( \frac{5\,\bnl}{12}\right)^2\,{\cal P}_\zeta\Nin\,
  \right]\,\ge\,0,
 \ee
 where $ \int_{\vec{q}}\,\equiv\,\int  d^3q/(2\pi)^3$.  
Provided that the quantity inside the square brackets is close to $k$ independent (for a discussion of this point in the context of the Suyama-Yamaguchi inequality \cite{Suyama:2007bg}, see \cite{Assassi:2012zq,Kehagias:2012pd}), we may conclude
with the following 
inequality
relating the variance of $\fnl$, with $\bnl$:
\be
\sigma_{\fnl}^2\,\ge\,\left( \frac{5\,\bnl}{12}\right)^2\,{\cal P}_\zeta\Nin.
\label{sigmafnl-general}\ee
This generalizes the inequalities we derived in the context of the local model to arbitrary shapes of non-Gaussianity.  For the special case of local non-Gaussianity, $\bnl$ depends on non-Gaussian parameters associated with four and three point functions, and mainly on $\gnl$ in set-ups in which there is a large hierarchy between $\gnl$, and $\taunl$ and $\fnl$\footnote{The left hand side of our general result (\ref{sigmafnl-general}) reduces to (\ref{sigmafnl}) and the right hand side to (\ref{sigmafnl-multi}) in the case of multi-source local non-Gaussianity, and neglecting all terms involving $\fnl$. The terms involving $\fnl$ and/or $\taunl$ are due to the variation of the power spectrum, an effect which we have neglected in this section. The fact that an appropriate squeezed limit of the 4-pt function mainly depends
on $\gnl$ was pointed out in  
 \cite{Pearson:2012ba}.}. In this limit, consistent with the previous sections, the  variance of $\fnl$ is proportional  to $\gnl^2$.
 
   Therefore our result is more general and does not only apply 
to the case of local non-Gaussianity, but also to other shapes that have interesting signals in the squeezed limit, for example the models recently discussed in \cite{Shiraishi:2013vja}.

\section{Conclusions}\label{sec:conclusions}

We have considered whether a large primordial trispectrum, characterised by the dimensionless parameter $\gnl$ \cite{Byrnes:2006vq}, is compatible with the stringent constraints on the value of the primordial bispectrum parameter, $\fnl$, allowed by Planck satellite data \cite{Ade:2013ydc}. 
Assuming the nearly scale-invariant distribution of primordial curvature perturbations extends for some range of scales beyond our observed horizon size, the long wavelength perturbations contribute to the effective background defined on smaller, observable patches. The contribution of these long wavelength fluctuations varies between different patches, making the explicit values of observables dependent upon the location of the patch.

We have shown that the probability of obtaining a small $\fnl$ in our horizon scale patch, for a model with a large value of $\gnl$ is small, regardless of the predicted value of $\fnl$ in the total inflated volume. The reason is that large $\gnl$ generates a variance of $\fnl$ between horizon-sized patches across the inflated volume. This means that even if a model naturally predicts $|\gnl|\gg\fnl^2$, it has to be tuned in order that $\gnl\sim10^5$ and $\fnl$ small enough to satisfy the observations. The model must additionally predict either the minimal possible number of efoldings consistent with our universe, otherwise we must be located in an atypical patch, in which $\fnl$ is smaller than in most other patches.

Our results are not dependent on any particular class of inflationary models. We assume only that primordial perturbations exist on a range of scales continuing beyond the observable scales within our horizon. For simplicity we have assumed they are almost scale-invariant for $\Nin$ e-foldings beyond our horizon, as might be expected in models of inflation which may well give rise to many more e-foldings than the $\sim60$ which are required to inflate our horizon scale. In the landscape picture of eternal inflation, huge inflating volumes may develop very different properties across a vast range of scales \cite{Linde:2005yw}. 
We restrict our analysis to the question of how correlators of the primordial curvature perturbations are related across a limited range of scales. 

For example, if $\gnl\gtrsim10^5$, and there are 30 additional e-foldings of approximately scale-invariant inflation before our Hubble horizon exited during inflation, then less than $10\%$ of the horizon sized patches will have a sufficiently small value of $\fnl$ to match observations, even if $\fnl=0$ globally. For $|\fnl|\gg10$ globally, the likelihood of observing small $\fnl$ is significantly smaller.

Our conclusions apply to local non-Gaussianity with single or multiple-sources. The variance of $\fnl$ grows as the number of relevant fields increases, therefore the most conservative bounds apply in the case of single-source models (those in which only one field contributes to the primordial curvature perturbation).
We have also shown that our conclusions apply to non-local shapes of non-Gaussianity evaluated in the appropriate squeezed limits. Our analysis reveals a new inequality between the 6-point function relating the variance of $\fnl$ between patches to the global 4-point function. This is reminiscent of the Suyama-Yamaguchi inequality, which relates the squeezed 3-point function ($\fnl$) to the collapsed 4-point function ($\taunl$), that has been shown to hold for any shape of non-Gaussianity. 

Although we have discussed super horizon patches, one may equally take the global values to be defined in our horizon and consider the variation between smaller patches within our horizon.  This may offer a way to test these results, for example with large scale structure surveys. Surveys within a finite patch may observe a value of $\fnl$ significantly different from the  average value over the whole CMB sky \cite{Byrnes:2011ri,Giddings:2011zd}.

\acknowledgments

We thank Mateja Gosenca, Elliot Nelson, Sarah Shandera and Tomo Takahashi for comments on a draft of this paper. CB and GT thank the Bethe Centre for Theoretical Physics, Bonn University, for hospitality where part of this work was completed. CB is supported by a Royal Society University Research Fellowship.
SN is supported by an Academy of Finland Grant number 257532. GT is
supported by an STFC Advanced Fellowship ST/H005498/1. DW is
supported by STFC grant ST/K0090X/1.

\end{document}